\documentclass[a4paper,12pt]{article}

\usepackage{hyperref}
\usepackage{amsmath,amsfonts,amssymb}
\usepackage{amsthm}
\usepackage{color}
\usepackage[normalem]{ulem}
\usepackage{cite}
\usepackage{graphicx}
\usepackage{comment}

\def\ds{\displaystyle}
\def\bea{\begin{array}{c}}
\def\ea{\end{array}}
\def\be{\begin{equation}\bea\ds}
\def\ee{\ea\end{equation}}
\def\bee{\begin{equation}\begin{array}{rcl}\ds}
\def\eee{\end{array}\end{equation}}

\def\tr{{\rm Tr}\,}

\def\tr{{\rm tr}\,}

\newcommand\Trule{\rule{0pt}{2.5ex}}
\newcommand\Brule{\rule[-1.ex]{0pt}{0pt}}

\title{Spectrum of Large N Glueballs: Holography vs Lattice }

\author{Anatoly Dymarsky$^a$ and Dmitry Melnikov$^b$}
\date{}

\begin{document}

\thispagestyle{empty}

\maketitle

\begin{center}
\textit{\small $^a$Department of Physics and Astronomy, \\ University of 
Kentucky, Lexington, KY, USA 40506}\\ \vspace{6pt}
\textit{\small  $^b$International Institute of Physics, Federal University of 
Rio Grande do Norte, \\ Campus Universit\'ario, Lagoa Nova, Natal-RN  
59078-970, Brazil}
\\ \vspace{2cm}
\end{center}

\vspace{-1cm}

\begin{abstract}
Recently there has been a notable progress in the study of glueball states in lattice gauge theories, in particular extrapolating their spectrum to the limit of large number of colors $N$. In this note we compare the large $N$ lattice results with the holographic predictions, focusing on the Klebanov-Strassler model. We note that glueball spectrum demonstrates approximate universality across a range of gauge theory models. Because of this universality the holographic models can give reliable predictions for the spectrum of pure  $SU(N)$ Yang-Mills theories with and without supersymmetry. This is especially important for the supersymmetric theories, for which no firm lattice predictions exist yet, and the holographic models remain the most tractable approach. For non-supersymmetric pure $SU(N)$ theories  with large $N$ we find an agreement within 5-8\% between the lattice and holographic predictions for the mass ratios of the lightest states in various sectors. In particular both lattice and holography give predictions for the  $2^{++}$ and $1^{--}$ mass ratio, consistent with the known constraints on the pomeron and odderon Regge trajectories.
\end{abstract}

\vspace{1cm}

In~\cite{TOTEM:2020zzr} the D0 and TOTEM collaborations announced a $3.4\sigma$ divergence of the $pp$ and $p\bar p$ cross sections, compatible with a $t$-channel exchange of a colorless $C$-odd particle, an ``odderon''~\cite{Lukaszuk:1973nt}. Further work is under way to improve the statistics of this result. The principle candidate for the odderon is a $C$-odd three-gluon bound state classified as $1^{--}$ in the $J^{PC}$ notations. The experimental evidence for odderon highlights importance of pure glue states, the glueballs.  Such states are important for understanding of the detailed structure of strong interactions~\cite{ParticleDataGroup:2020ssz} as well as possible extensions of physics beyond the  Standard Model, see e.g.~\cite{Kang:2008ea,Juknevich:2009ji,Juknevich:2009gg}. 

Existence of pure glue bound states was conjectured long ago, but their experimental detection, and elucidating their physical properties, turned out to be a complicated task.  The main theoretical challenge is  the heavy mixing of such nonquarkyonic short-lived states with heavy excited states of ordinary mesons. As a result  most of the presently available data concerning glueballs come from the lattice simulations. Mixing with quarks is a challenge for the lattice as well, so most of the time the studies assume the quenched approximation, effectively studying the pure glue (or pure Yang-Mills) sector of QCD.

The standard reference for the spectrum of glueballs is the work of Morningstar and Peardon~\cite{Morningstar:1999rf}, which found 12 lightest  states with spins varying from $J=0$ to $J=3$ in the pure glue $SU(3)$ Yang-Mills theory.\footnote{See~\cite{Michael:1988jr,Michael:1989ry,Bali:1993fb} for yet earlier work on the lattice spectrum.} The ensuing lattice studies improved the original predictions of the glueball masses~\cite{Chen:2005mg,Meyer:2004gx,Athenodorou:2020ani}, observing additional excited states, studied the effects of introducing quarks~\cite{Bali:2000vr,Hart:2001fp,Hart:2006ps,Richards:2010ck,Gregory:2012hu,Sun:2017ipk,Brett:2019tzr,Chen:2021dvn}, investigated  dependence on the gauge group and its rank~\cite{Teper:1998kw,Lucini:2004my,Lucini:2010nv,Holligan:2019lma,Bennett:2020hqd,Bennett:2020qtj,Athenodorou:2021qvs} and attempted to estimate the decay constants of a few lightest states~\cite{Chen:2005mg}, cf.~\cite{Sexton:1994wg,Sexton:1996ed,Yamanaka:2019yek,Llanes-Estrada:2021evz}. 

The  glueball spectrum exhibits relatively mild dependence on the number of colors, in  the 't Hooft limit\footnote{In the $Sp(N)$ case the leading correction to the $N\to\infty $ value is $O(1/N)$.}
\be
\label{MassNc}
m \ \simeq \ m_\infty + \frac{c}{N^2}+\dots
\ee
expected from a large $N$  expansion and corroborated by the lattice analysis \cite{Lucini:2001ej,Lucini:2012gg,Lucini:2013qja}.
In table~\ref{tab:SUN} we show lattice results for $SU(N)$~\cite{Lucini:2004my} and $Sp(N)$~\cite{Bennett:2020qtj}  for the mass of the lightest $0^{++}$ glueball and its ratio with the mass of the lightest excitation, the $2^{++}$ state. For $SU(N)$ with $N$ ranging from $N=2$ to $N=\infty$ the  variation of masses is within 14-16\%, while the variation of the ratios is even smaller, about 10\%. In the $Sp(N)$ case the variation of the ratios is even smaller, about 5\%.
\begin{table}[h]
    \centering
    \begin{tabular}{||c|cccccc||}
      \hline \Trule\Brule $G$  & $SU(2)$ & $SU(3)$ & $SU(4)$ & $SU(6)$ & $SU(8)$ & $SU(\infty)$ \\
        \hline\Trule\Brule $m_{0^{++}}$  & 3.78 & 3.55 & 3.36 & 3.25 & 3.55 & 3.31 \\ 
        \Trule\Brule$m_{2^{++}}/m_{0^{++}}$ & 1.44 & 1.35 & 1.45 & 1.46 & 1.32 & 1.46 \\
        \hline\hline \Trule\Brule $G$ & $Sp(1)$ & $Sp(2)$ & $Sp(3)$ & $Sp(4)$ & -- & $Sp(\infty)$ \\
        \hline \Trule\Brule $m_{2^{++}}/m_{0^{++}}$ & 1.41 & 1.41 & 1.48 & 1.41 & -- & 1.47 \\ \hline
    \end{tabular}
    \caption{Lattice masses (in QCD string tension units~\cite{Bennett:2020qtj}) and mass ratios of the $0^{++}$ and $2^{++}$ glueballs in the pure glue $SU(N)$~\cite{Lucini:2004my} and $Sp(N)$ theories~\cite{Bennett:2020qtj}.}
    \label{tab:SUN}
\end{table}

Another important observation is a reasonably small effect of quark mixing in the unquenched version of the Yang-Mills theory. The phenomenological OZI rule~\cite{Okubo:1963fa,Zweig:1964jf,Iizuka:1966fk} does not favor quark mixing with purely gluonic states. Lattice simulations confirm this effect to certain extent, as can be seen from the data in table~\ref{tab:unquenched}. The match between quenched and unquenched cases  is better for the lighter states (5\% for $0^{++}$), and the ratio $m_{2^{++}}/m_{0^{++}}\sim 1.46$ obtained for QCD with three flavors is compatible with the pure  glue $SU(N)$ and $Sp(N)$ results~(approximate universality of this ratio was also acknowledged in~\cite{Bennett:2020hqd}). Mass ratios appears to be even more stable with respect to model variation, which might indicate that not only $m_\infty$, but also $c$ in equation~(\ref{MassNc}) are universal.

\begin{table}[h]
    \centering
    \begin{tabular}{||c|cccc||ccc||}
        \hline \Trule\Brule $J^{PC}$ & $0^{++}$ & $2^{++}$ & $1^{+-}$ & $0^{-+}$ & $m_{2^{++}}/m_{0^{++}}$ & $m_{1^{+-}}/m_{2^{++}}$ & $m_{0^{-+}}/m_{2^{++}}$ \\
        \hline \Trule\Brule  YM & 1710 & 2390 & 2980 & 3640 & 1.40 & 1.25 & 1.52 \\
        \Trule\Brule QCD$_3$ & 1795 & 2620 & 3270 & 4490 & 1.46 & 1.25 & 1.71 \\ \hline
    \end{tabular}
    \caption{Comparison of the lattice glueball masses (measured in MeV) in the quenched Yang-Mills and QCD with three flavors~\cite{Gregory:2012hu}. The ratio of the masses of $m_{2^{++}}$ and $m_{0^{++}}$ for the two cases are $1.40$ and $1.46$ respectively.}
    \label{tab:unquenched}
\end{table}

Approximate universality of the glueball spectrum,  in the sense of weak $N$-dependence, makes it natural to study it using holographic models. Classical gravity limit of holography~\cite{Maldacena:1997re,Gubser:1998bc,Witten:1998qj}, the so-called gauge/gravity duality, can only capture the limit of large number of colors $N\to\infty$ and large 't Hooft coupling constant $\lambda=g_{YM}^2N\to\infty$ of Yang-Mills theory. Without taking into account quantum corrections to gravity one can only hope to access some universal large $N$ properties of hadrons. Lattice simulations indicate that the glueball spectrum, which shows approximate $N$ independence, is one of such quantities which can be accessed holographically.

Sparing the details, the holographic calculation in the classical gravity limit boils to choosing a background -- a solution of the appropriate (super)gravity equations -- that is  dual  to the gauge theory of interest, and calculating the spectrum of linearized excitations by expanding equations of motion to linear order above the background solution. The spectrum of gravity excitations is then equivalent to the spectrum of gauge theory. Higher order correlation functions are also available though require calculations beyond the linear order.

Early take on glueballs in holography~\cite{Csaki:1998qr,Brower:2000rp} was in the case of the so-called Witten's model~\cite{Witten:1998zw}. The background of this model is the near-horizon limit of the world-volume theory of D4 branes compactified on a circle. This background preserves no supersymmetry and can be used to  qualitatively describe gauge theory spectrum. As a drawback, it exhibits features incompatible with the Yang-Mills theory, namely mass degeneracy of different glueball families, which points to additional symmetries beyond those of the Yang-Mills theory. It also exhibits larger spacing of masses, as demonstrated in table~\ref{tab:holomodels}. In Witten's model the ratio $m_{2^{++}}/m_{0^{++}}\sim 1.74$, significantly larger than in the pure $SU(N)$ case. The discrepancy is even bigger for heavier states.

\begin{table}[h]
    \centering
   \begin{tabular}{||c|ccccc||}
      \hline \Trule\Brule $J^{\,PC}$ & $SU(\infty)$ & $Sp(\infty)$ & BMT & BBC$_D$ & BBC$_N$ \\
      \hline
      \Trule\Brule ${0^{++}}$ & 1 & 1 & 1 & 1 & 1 \\
      \Trule\Brule ${2^{++}}$ & 1.49 & 1.47 & 1.74 & 1.48 & 1.56\\
      \Trule\Brule ${0^{-+}}$ & 1.53 & 1.54 & 2.09 & -- & --\\
      \Trule\Brule $1^{+-}$ & 1.88 & --  & 2.70 & -- & -- \\
      \Trule\Brule $1^{--}$ & 2.32 & --  & 3.37 & -- & -- \\
      \Trule\Brule $0^{+-}$ & 3.01 & --  & -- & -- & -- \\
      \hline\hline
      \Trule\Brule $0^{++\ast}$ & 1.89 & 1.94 & 2.53 & 1.63 & 1.83 \\
      \Trule\Brule $2^{++\ast}$ & 2.11 & -- & 2.76 & 2.15 & 2.49 \\
      \hline
    \end{tabular}	
    \caption{Spectra of the lightest glueballs in Witten's (BMT~\cite{Brower:2000rp}) and hard wall models (BBC~\cite{BoschiFilho:2005yh}) compared to the state of the art lattice extrapolations of $m_\infty$ for $SU(N)$~\cite{Athenodorou:2021qvs} and $Sp(N)$~\cite{Bennett:2020qtj}. The masses are normalized to the mass of the $0^{++}$ state. Models BBC$_N$ and BBC$_D$ used different boundary conditions (Neumann or Dirichlet) in the calculation of the spectrum.}
    \label{tab:holomodels}
\end{table}

A straightforward approach to model hadron physics holographically can be taken via bottom-up models, the so-called hard wall model would be the simplest case~\cite{Erlich:2005qh,Polchinski:2001tt,Polchinski:2002jw}. For the purpose of this note, the Light-Front Holography models~\cite{Brodsky:2006uqa} can also be attributed to this class. In the hard wall model the background is five-dimensional anti de Sitter space ($AdS_5$), whose group of symmetries is precisely the conformal group in 3+1 dimensions, and a cutoff is introduced to break this symmetry explicitly. The dual theory to such a background at high energies would qualitatively be a (3+1)-dimensional conformal  gauge theory, with an IR scale defining the masses, as in pure Yang-Mills or QCD. One then considers wave equations  for matter fields of various spin in the $AdS_5$ background with the  boundary conditions at the cut off radius, i.e.~at the hard wall. This mathematical problem leads to the spectrum of light states, as calculated by~\cite{BoschiFilho:2005yh} and shown in table~\ref{tab:holomodels}.

Remarkably, the plain hard wall model predicts the value $m_{2^{++}}/m_{0^{++}}\sim 1.48$,  that is very close to the extrapolated ratio of $m_\infty$ calculated on the lattice. This ``magic number" can be expressed as a ratio of the first non-trivial zeroes $x_{2,1}$ and $x_{4,1}$ of Bessel functions $J_2(x)$ and $J_4(x)$,
\be
\text{hard wall:}\quad \frac{m_{2^{++}}}{m_{0^{++}}} \ = \ \frac{x_{4,1}}{x_{2,1}} \ \simeq \ 1.47759\,.
\ee 
The hard-wall model also has its drawbacks. In particular, it does not include states with non-trivial parity and charge conjugation, since the model does not have a natural way to implement corresponding symmetries. In contrast, top-down holographic constructions inherit these symmetries from string theory, as in the case of Witten's model. 

The discussion above, which loosely follows historic developments, suggests that a more nuanced top-down model would be necessary to accurately describe large $N$ physics  in pure Yang-Mills and QCD-like theories. The Klebanov-Strassler (KS) model~\cite{Klebanov:2000hb} was proposed with this goal in mind. It is a top-down holographic model, which possesses rich IR physics. It is a low-energy limit of D3 and D5 branes in type IIB string theory compactified on a six-dimensional cone (conifold). The compactification gives rise to a type IIB supergravity background on a space which is a warped product of $AdS_5$ and a pair of spheres $S^3\times S^2$~\cite{Klebanov:1998hh,Klebanov:1999rd,Klebanov:2000nc}. Smoothing out the singularity of the cone (deformation of the conifold) introduces an IR scale that breaks (3+1)-dimensional conformal invariance and provides an interesting example of a holographic RG flow with a logarithmic running of the coupling constants. The background has $SU(2)\times SU(2)\times U(1)_{\rm B}$ global symmetries, so its spectrum is organized in the irreducible representations of this group.

The Klebanov-Strassler background preserves ${\cal N}=1$ supersymmetry, so the dual gauge theory is a non-conformal ${\cal N}=1$ supersymmetric Yang-Mills theory with additional matter fields transforming under the global symmetry group, in the presence of a particular superpotential. In the IR the theory flows to a strongly coupled fixed point (the corresponding RG flow is known as a ``cascade'' of Seiberg dualities~\cite{Klebanov:2000hb,Strassler:2005qs}) which was initially thought to be in the universality class of the pure supersymmetric $SU(N)$ Yang-Mills theory. It was then understood that the IR limit of the Klebanov-Strassler theory has additional massless states due to spontaneous breaking of the baryon $U(1)_{\rm B}$ symmetry by the baryonic operators~\cite{Gubser:2004qj,Gubser:2004tf,Benna:2006ib}. 

As a close relative of the pure ${\cal N}=1$ Yang-Mills, the Klebanov-Strassler gauge theory shares with it many core properties. Apart from the mentioned logarithmic running of the coupling constants, the holographic theory exhibits the same mechanism of breaking of the $U(1)$ R-symmetry~\cite{Klebanov:2002gr}. The singlet sector mostly contains the super Yang-Mills operators~\cite{Apreda:2003gc,Cassani:2010na}. In view of the approximate universality of the spectrum, one can hope that $SU(2)\times SU(2)$-singlet states  reproduce the appropriate limit of the low-energy spectrum of the supersymmetric, or even pure bosonic $SU(N)$ Yang-Mills. With this logic in mind, we summarize below the results for the singlet states masses in the Klebanov-Strassler theory.

The spectrum of singlet fluctuations of the Klebanov-Strassler background was studied in a series of papers including~\cite{Amador:2004pz,Berg:2005pd,Dymarsky:2006hn,Berg:2006xy,Dymarsky:2007zs,Benna:2007mb,Dymarsky:2008wd,Gordeli:2009nw,Gordeli:2013jea,Melnikov:2020cxe}. (See also~\cite{Caceres:2005yx,Elander:2009bm,Bianchi:2010cy,Elander:2014ola,Elander:2017cle,Elander:2017hyr} for the studies of non-singlet states or deformations of the Klebanov-Strassler theory.) The structure of the spectrum in the supergravity limit was explained in~\cite{Gordeli:2009nw, Gordeli:2013jea,Melnikov:2020cxe}: it contains two massless scalar supermultiplets and thirteen massive supermultiples including one graviton, two gravitino, four vector and six scalar supermultiplets. The latter are massive ${\cal N}=1$ supermultiplets labeled by the highest spin component, except for the scalar multiplet, which  is conventionally labeled by its lowest spin component. 

\begin{figure}
    \centering
    \includegraphics[width=0.45\linewidth]{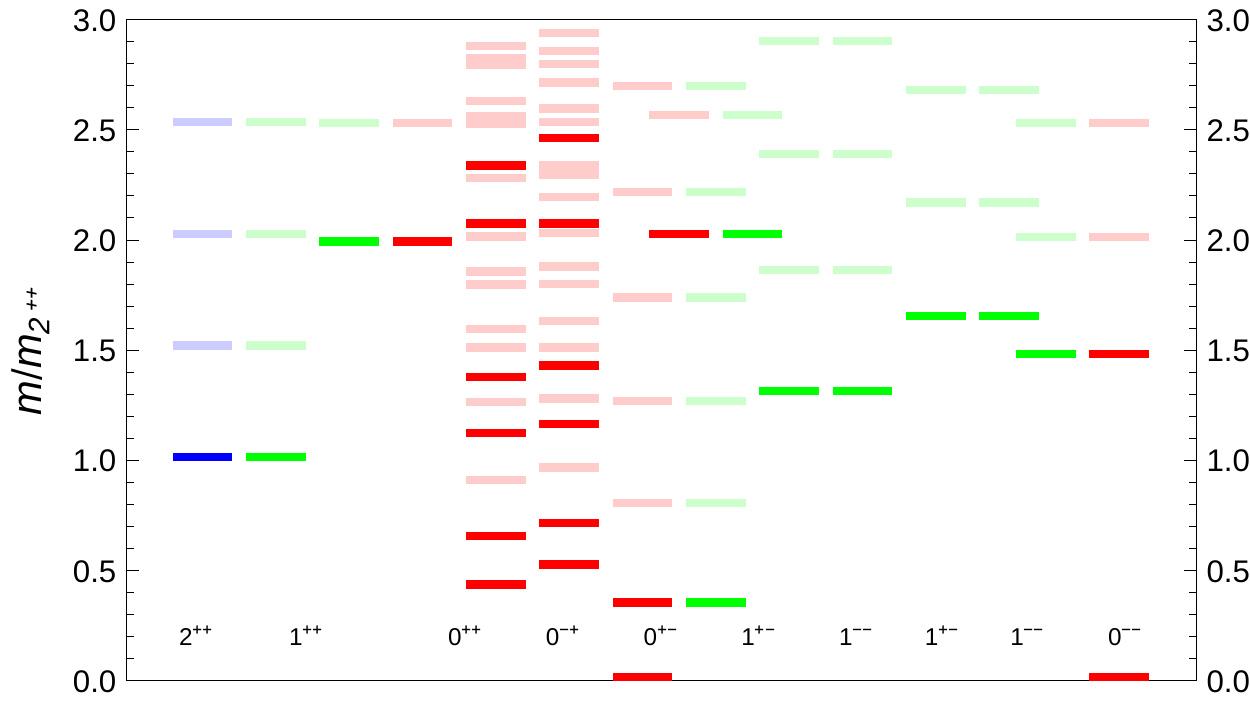} \quad
    \includegraphics[width=0.45\linewidth]{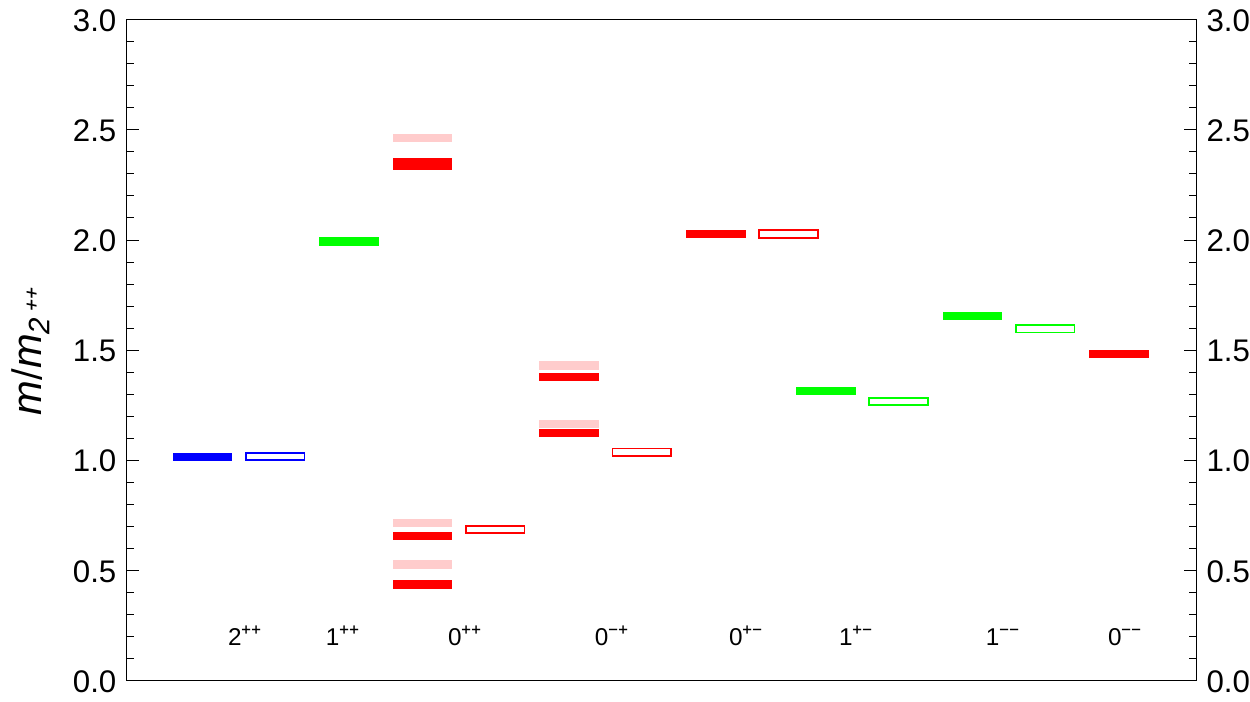}
    \caption{Full light $SU(2)\times SU(2)$-singlet spectrum of the bosonic states in the Klebanov-Strassler theory (left). The ground states of each supermultiplet are shown in full color, while the excited states are faded. All the mass values are computed in the units of the ground state of $2^{++}$. The discrepancy of the $0^{++}$~\cite{Berg:2006xy} and $0^{-+}$~\cite{Melnikov:2020cxe} sectors is visible. The right plot compares the masses of the ground states in the supersymmetric Yang-Mills sector (filled rectangles) and the ground states in $SU(N)$ bosonic Yang-Mills, extrapolated to $N=\infty$ (empty rectangles). In the $0^{++}/0^{-+}$ sector the results of~\cite{Berg:2006xy} are shown in full color, while that of~\cite{Melnikov:2020cxe} are faded. States that do not have pairs correspond to hybrid glueballs with no analog in the bosonic theory, except for the heaviest $0^{++}$ that has not been approached on the lattice.}
    \label{fig:fullspec}
\end{figure}

 Our goal is to compare this spectrum of Klebanov-Strassler theory with the pure gauge ${\cal N}=1$ Yang-Mills. Since the KS theory exhibits additional global symmetries only certain states (multiplets) are the counterparts of the pure YM. We elaborate this point below. 
The massless states of the KS theory are not part of the Yang-Mills spectrum. 
One should also exclude multiplet coming from the $U(1)_{\rm B}$ sector: a scalar 
multiplet as well as a vector multiplet containing $0^{+-}$ and $1^{+-}$ 
~\cite{Benna:2007mb,Dymarsky:2008wd}. The remaining eleven massive multiplets 
exist in the spectrum of the pure supersymmetric Yang-Mills. They are listed in 
table ~\ref{tab:KSsinglets} and their spectrum is illustrated by 
figure~\ref{fig:fullspec}.

\begin{table}[h]
    \centering
    \begin{tabular}{||c|c|c|cc|c||}
     \hline \Trule\Brule & $J^{\,PC}$ & Multiplet & $m^{\rm GS}$ & $m^{\ast}$  & Ref.\\
      \hline
      \Trule\Brule 1 & $1^{++},2^{++}$ & graviton & 1 & 1.51  & \cite{Dymarsky:2006hn}\\ \hline
      \Trule\Brule 2 & $1^{+-},1^{--}$ & gravitino & 1.30 & 1.85  & \cite{Dymarsky:2008wd} \\ 
      \Trule\Brule 3 & $1^{+-},1^{--}$ & gravitino & 1.64 & 2.15  & \cite{Dymarsky:2008wd} \\ \hline
      \Trule\Brule 4 & $0^{--},1^{--}$ & vector & 1.47 & 2.00  & \cite{Benna:2007mb} \\ 
       \Trule\Brule 5 & $0^{+-},1^{+-}$ & vector & 2.01 & 2.55  & \cite{Benna:2007mb} \\ \hline
       \Trule\Brule 6 & $0^{++},1^{++}$ & vector & 1.99 & 2.53  & \cite{Gordeli:2009nw} \\ \hline\hline
       \Trule\Brule 7 & $0^{++},0^{-+}$ & scalar & 0.421 & 0.894  & \cite{Berg:2006xy} \\ \hline
       \Trule\Brule 8 & $0^{++},0^{-+}$ & scalar & 0.640 & 1.25  & \cite{Berg:2006xy} \\ 
       \Trule\Brule 9 & $0^{++},0^{-+}$ & scalar & 1.11 & 1.58  & \cite{Berg:2006xy} \\ \hline\hline
        \Trule\Brule 10 & $0^{++},0^{-+}$ & scalar & 1.36 & 1.84  & \cite{Berg:2006xy} \\ 
         \Trule\Brule 11 & $0^{++},0^{-+}$ & scalar & 2.32 & 2.87  & \cite{Berg:2006xy} \\ \hline
    \end{tabular} \quad
    \begin{tabular}{||c|cc||}
     \hline \Trule\Brule $J^{\,PC}$  & $m^{\rm GS}$ & $m^{\ast}$  \\
      \hline
      \Trule\Brule $2^{++}$  & 1 & 1.43  \\ \hline
      \Trule\Brule $1^{+-}$  & 1.25 & 1.62   \\ 
      \Trule\Brule $1^{--}$  & 1.58 & $\geq 1.88$  \\ \hline
      \Trule\Brule    &  &       \\ 
      \Trule\Brule $0^{+-}$  & $\geq2.01$ & --  \\ \hline
      \Trule\Brule    &  &       \\  \hline\hline
      \Trule\Brule   &  &    \\  \hline
       \Trule\Brule $0^{++}$  & 0.668 & 1.27  \\ 
       \Trule\Brule $0^{-+}$  & 1.02 & 1.53  \\ \hline\hline
       \Trule\Brule   &  &    \\  
       \Trule\Brule  $0^{++}$ & -- & --   \\  \hline
    \end{tabular}
    \caption{Spectrum of the lightest supermultiplets of the supersymmetric Yang-Mills sector of the Klebanov-Strassler theory (left table). The bosonic members of the multiplets are indicated. The masses of the ground and first excited states, in units of the mass of the ground state of $2^{++}$, are shown, as well as the references to the works, from which the values are  extracted. For the scalar sector, the assignment was made based on the analysis of~\cite{Melnikov:2020cxe,RodriguesFilho:2020rae}. These masses are compared with the masses recently obtained for the $SU(\infty)$ bosonic Yang-Mills on the lattice~\cite{Athenodorou:2021qvs} (right table). Lower bounds indicate states with known masses, but with a problem of confirming the $J^{PC}$ numbers in the continuum limit.}
    \label{tab:KSsinglets}
\end{table}

In table~\ref{tab:KSsinglets} we measure all  masses  in the units of the mass of the ground state of the graviton multiplet, the $2^{++}$ glueball. This state is particularly easy to analyze on the supergravity side,  since the $2^{++}$ fluctuation above the background decouples from all the other modes~\cite{Dymarsky:2006hn,Bachas:2011xa}. It thus serves  a natural reference point for the remaining states. The charge conjugation-odd ($C$-odd) sector is also relatively simple to analyze holographically. The masses of the $1^{+-}$, $1^{--}$ and $0^{+-}$ glueballs can be compared with the $SU(N)$ bosonic Yang-Mills extrapolated to $N=\infty$~\cite{Athenodorou:2021qvs}. We note  the holographic masses have the same  hierarchy as on the lattice. Even more, they match the $SU(\infty)$ ground state estimates within 5\%. For the excited states the correspondence is not so good: the holographic supersymmetric masses of excited states grow more rapidly with the excitation number than their lattice bosonic counterparts.

Masses of some of the states calculated on the lattice in~\cite{Athenodorou:2021qvs} and shown in table~\ref{tab:KSsinglets}, appear as lower bounds, as for the $1^{--\ast}$ and $0^{+-}$ states. This is because the unambiguous identification of the $J^{PC}$ quantum numbers of these states is not yet possible on the lattice, and one can only classify them tentatively  as $1^{--\ast}$ and $0^{+-}$, based on general mass hierarchy arguments. We note that in this case the holographic results support this identification. 

Not all the multiplets of the supersymmetric Yang-Mills theory survive in the quenched approximation. Some of the multiplets only contain composite states of fermion operators. The two examples are vector multiplets 4 and 6 in table~\ref{tab:KSsinglets}. For such states there are no lattice results yet to compare.

The most complex is the scalar sector, originally studied in~\cite{Berg:2005pd,Berg:2006xy}. There are six pairs of scalar (and also six pseudoscalar) multiplets which mix together. The scalars contain the purely bosonic $0^{++}$ and $0^{-+}$ modes and scalars associated with the composite gluino operators. The scalar multiplet containing the gluino bilinear $\lambda\lambda$ state is expected to be the lightest in the spectrum of the supersymmetric Yang-Mills. Apart from $\tr\lambda\lambda$ and the scalars due to $\tr F_{\mu\mu}F^{\mu\nu}$ and $\tr F_{\mu\nu}\tilde{F}^{\mu\nu}$ operators, the  holographic model  includes other $0^{++}$ and $0^{-+}$  gluon-gluino states associated with the composite operators of dimension $\Delta=3,4,5,6,7,8$ (see~\cite{Apreda:2003gc,Cassani:2010na,Gordeli:2013jea,Melnikov:2020cxe} for the classification).

The  holographic spectrum of $0^{++}$ states was originally calculated in~\cite{Berg:2006xy}. The result of that calculation is a list of lowest masses, which on physical grounds are expected to belong to independent towers associated with seven individual particles and their excitations. In the conformal case these would be related to seven composite operators discussed in ~\cite{Apreda:2003gc,Cassani:2010na}. It is not clear how to split excitations into towers from the first principles. That is except for one tower associated with a particle from the vector supermultiplet. It can be separated by comparison with its  superpartner vector mode, which decouples from all other modes on gravity side, see~\cite{Gordeli:2009nw}. For all other states, below  we use that masses approximately follow ``Regge trajectories'' associated with the radial quantum number $n$, and identify ground states for each tower.

 In~\cite{Melnikov:2020cxe} the spectrum of $0^{-+}$, the superpartners of $0^{++}$,   was calculated as a consistency check. It was proposed that the following ordering of states is consistent with \cite{Berg:2006xy,Melnikov:2020cxe} and with the lattice results~\cite{RodriguesFilho:2020rae},
\be
m_{\lambda\lambda} < m_{0^{++}} < m_{\lambda\lambda}^* < m_{0^{-+}} < m_{0^{++}}^* < \cdots\,, \label{tri}
\ee
where $m_{\lambda\lambda}$ is the mass of the ground state of the gluino bilinear, $m_{\lambda\lambda}^\ast$ is its first excited state, $m_{0^{++}}$ and $m_{0^{-+}}$ are masses of the states in the spectrum of the two gluon operators $\tr F_{\mu\nu}F^{\mu\nu}$ and $\tr F_{\mu\nu}\tilde{F}^{\mu\nu}$. Unlike gluino operators, which pose a challenge for lattice methods,  the latter operators are commonly studied on the lattice. 

\begin{figure}
    \centering
    \includegraphics[width=0.45\linewidth]{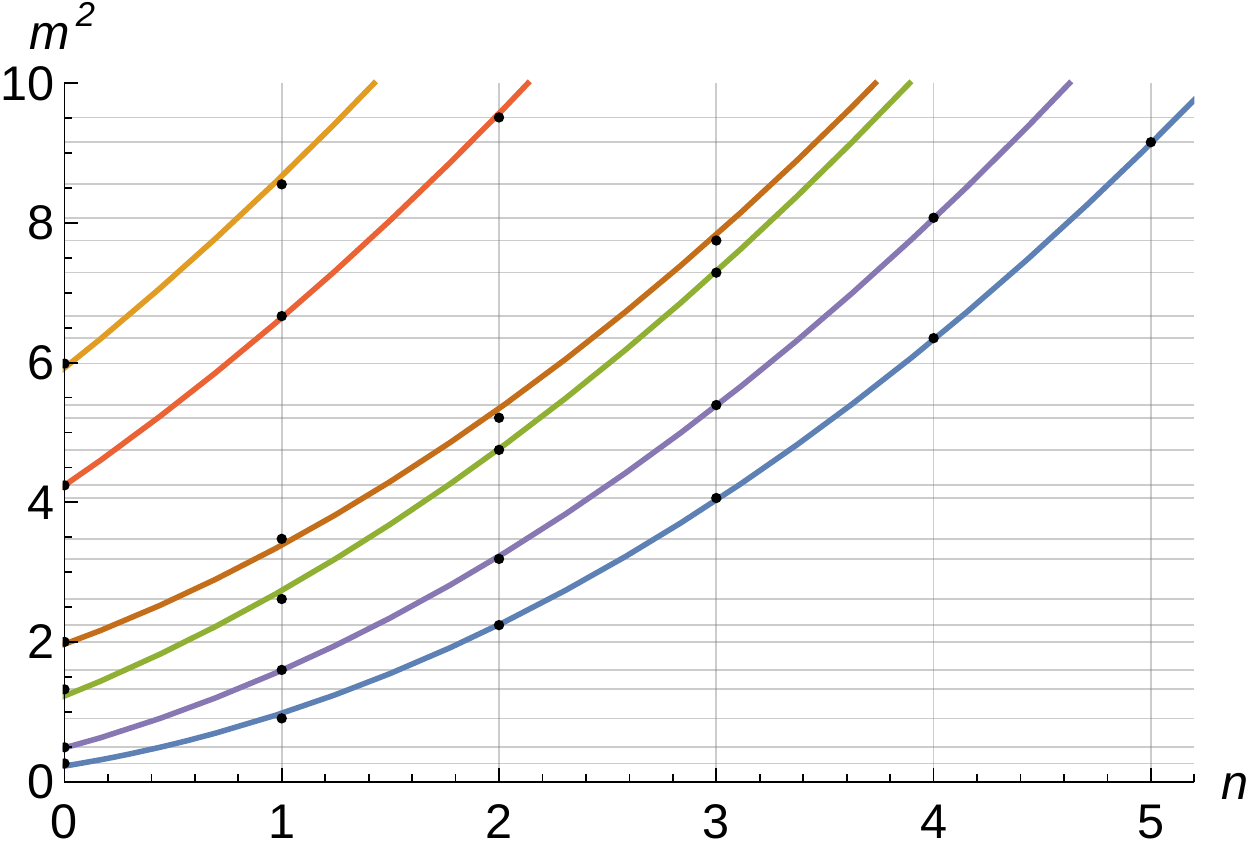}
    \quad
    \includegraphics[width=0.45\linewidth]{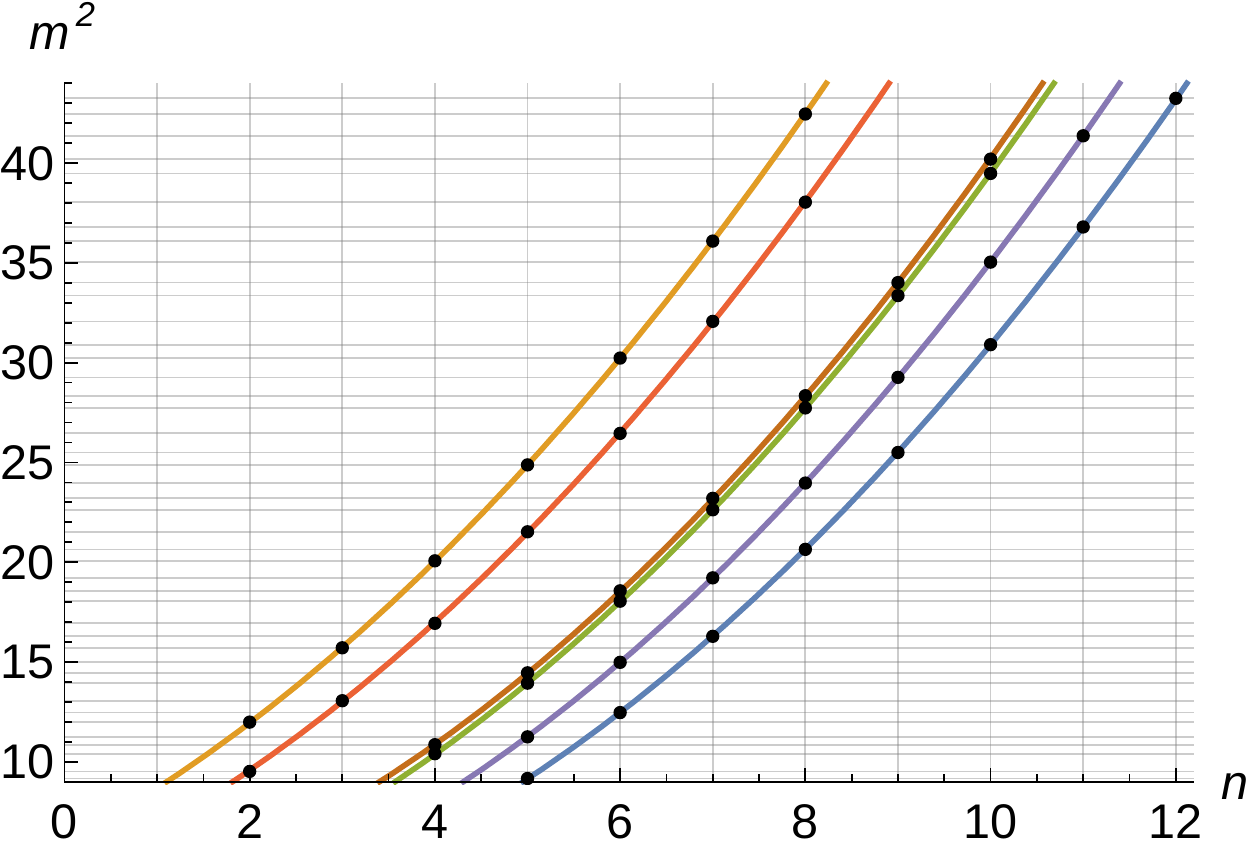}
    \caption{Quadratic fits of the $0^{-+}$ spectrum of $m^2$ found in~\cite{Melnikov:2020cxe}. The black dots (as well as the grids) indicate the masses and the curves are fits~(\ref{mll})-(\ref{m11}). $n$ is the ``radial'' quantum number labeling excited states.
The left panel shows the behavior of the fits for $m^2\leq 10$. The intercepts correspond to the ground states. The right panel illustrates the quality of the fits for $m^2\geq 10$. All mass values are shown in the units of the ground state mass of $2^{++}$.}
    \label{fig:fits}
\end{figure}

In the present note we extend \eqref{tri} and complete the hierarchy of  masses by adding  two remaining multiplets. One of these correspond to four-gluon operator $\tr (F_{\mu\nu}F^{\mu\nu})^2$, which is in principle accessible on the lattice. This is done by applying quadratic fit to the values of masses squared. This approach worked very well in the past for other sectors of the KS theory~\cite{Dymarsky:2006hn,Dymarsky:2007zs,Benna:2007mb,Dymarsky:2008wd,Gordeli:2013jea}. The results of the fitting of the pseudoscalar sector~\cite{Melnikov:2020cxe} are shown in figure~\ref{fig:fits}. The following fits work very well for the heavy part of the spectrum, while for the light states we end up with a few noticeable deviations,
\begin{eqnarray}
m_{\lambda\lambda}^2 \ = \ 0.223 +0.499n + 0.257 n^2 \,, &&  m^{\rm GS} \ = \ 0.472\,, \label{mll}\\
m_{0^{++}}^2 \ = \ 0.480 + 0.854 n + 0.260 n^2\,, & &  m^{\rm GS} \ = \ 0.693\,,\\
m_{0{-+}}^2 \ = \ 1.22 + 1.26 n + 0.257 n^2\,, & & m^{\rm GS} \ = \ 1.11 \,, \\
m_{10}^2 \ = \ 1.97 + 1.15 n + 0.267 n^2\,, & & m^{\rm GS} \ = \ 1.40 \,, \\
m_{AB}^2 \ = \ 4.24 + 2.14 n + 0.261 n^2\,, & & m^{\rm GS} \ = \ 2.06 \,, \\
m_{11}^2 \ = \ 5.92 + 2.49 n + 0.260 n^2\,, & & m^{\rm GS} \ = \ 2.43 \,. \label{m11}
\end{eqnarray}
Here  all the values are given in units of the mass of the ground state of $2^{++}$ and $m^{\rm GS}$ stands for $n=0$, i.e.~extrapolated ground state value of the corresponding tower. Here $m_{10}$ and $m_{11}$ are pseudoscalar predictions for the masses of the ground states in the entries 10 and 11 of table~\ref{tab:KSsinglets}. Masses $m_{AB}$ cannot be compared with the spectrum of the pure supersymmetric Yang-Mills, as the corresponding operators come from the $U(1)_{\rm B}$ sector and would be absent in pure glue theory. We note that using $m^{\rm GS}=m(n=0)$ rather than actual lowest mass slightly improves the convergence to the $SU(\infty)$ lattice values.

It is necessary to mention that the spectrum of pseudoscalars computed in~\cite{Melnikov:2020cxe}, although similar  with the spectrum of $0^{++}$ from~\cite{Berg:2006xy}, and features the same hierarchy and large $n$ asymptotic,  exhibits a consistent divergence of the masses of two out of six towers. This divergence is particularly pronounced for low masses. For example, in~\cite{Berg:2006xy} the masses of the three lightest scalar ground states are $m_{\lambda\lambda}\simeq 0.511\,m_{2^{++}}$, $m_{0^{++}}\simeq 0.701\,m_{2^{++}}$ and $m_{0^{-+}}\simeq 1.15\,m_{2^{++}}$, cf. (\ref{mll})-(\ref{m11}), to be compared with the entries 7, 8 and 9 in table~\ref{tab:KSsinglets}. See also figure~\ref{fig:fullspec}. The spectrum of~\cite{Berg:2006xy} was independently confirmed in~\cite{Elander:2014ola}, so the status of the pseudoscalar calculation remains unclear.

\bigskip

We conclude by discussing relevance of holographic results in the context of lattice simulations. Recent progress on the lattice largely confirms the expectation of universality  -- approximate $N$ independence in the sense of \eqref{MassNc} -- of the glueball masses. This allows to extrapolate the spectrum to the limit of infinite number of colors, making possible  comparison of the lattice results for $SU(\infty)$ with the Yang-Mills subsector of the Klebanov-Strassler theory.  Before proceeding, let us stress a subtlety important for any comparison of this kind. 

The models conventionally studied on the lattice are asymptotically free and have no tunable parameters except for dimensionful $\Lambda_{QCD}$, or equivalently the confining string tension $\sigma$. Hence, the spectrum on the lattice can be unambiguously expressed in units of $\sqrt{\sigma}$. Meanwhile, in the KS theory, as well as other models of gauge/(super)gravity duality, there is a free parameter, an exactly marginal coupling associated with the value of the dilaton. On the dual field theory side this is the 't Hooft coupling $\lambda$.\footnote{In the KS theory, in addition to an exactly marginal $\lambda$ there is also  a running coupling.} Accordingly in holographic models the ratio $m_{J^{PC}}/\sqrt{\sigma}$ is $\lambda$-dependent, making direct comparison with the lattice problematic. Furthermore in the strong coupling limit $\lambda\gg 1$, when gravity approximation is valid, $m_{J^{PC}}/\sqrt{\sigma}$ is suppressed by a positive power of $\lambda$, see~\cite{Klebanov:2000hb,Dymarsky:2005xt}. Strictly speaking, the mass-tension ratios  vanish in the supergravity limit. However, for glueballs of small spin  the leading scaling of $m_{J^{PC}}/\sqrt{\sigma}$ with $\lambda$ is universal, rendering glueball mass ratios $\lambda$-independent. They can be compared directly with the lattice results, as discussed below.

For the higher spin states with $J\geq 2$, except for the $2^{++}$ which behaves as $J<2$ states, in the strong coupling limit, the $\lambda$-dependence is different. In fact these masses diverge in supergravity limit, as they belong to massive string sector and their masses are of order $1/\sqrt{\alpha'}$. In other words these states are beyond (super)gravity approximation, making comparison with the lattice presently impossible.  

With this said we  proceed by comparing the holographic mass ratios  with those on the lattice. For the five lightest states with $J<2$ and for the $2^{++}$ state of the bosonic Yang-Mills theory we observed the same hierarchy of the spectrum in both the Klebanov-Strassler theory and on the lattice. Furthermore,  for the five mass ratios of these six lightest states we observe a better than ten percent consistency of the numerical values, as summarized in table~\ref{tab:KSratios} for both $SU(\infty)$ and $SU(3)$ theories. These results provide an additional evidence for the universality of the spectrum, and an independent check for the consistency of the lattice approach.

\begin{table}[h]
    \centering
    \begin{tabular}{||c|c|ccccc||}
\hline \Trule\Brule $J^{\,PC}$ & $2^{++}$ & $0^{++}$ & $0^{-+}$ &  $1^{+-}$ &  $1^{--}$ &  $0^{+-}$\\
      \hline
      \Trule\Brule Holography/$SU(\infty)$ & 1 & 0.959 & 1.081 & 1.037 & 1.038 & 0.999 \\ 
      \Trule\Brule Holography/$SU(3)$ & 1 & 0.920 & 1.027 & 1.048 & 0.967 & 1.057 \\ \hline
    \end{tabular}
    \caption{Ratios of glueball masses measured in the units of the $2^{++}$ mass, $({m_{\rm hol}^{J^{PC}}/m_{\rm hol}^{2^{++}})/( m_{\rm latt}^{J^{PC}} }/m_{\rm latt}^{2^{++}})$, for holographic (Klebanov-Strassler) and lattice predictions for the lightest glueball ground states in the $SU(\infty)$ and $SU(3)$ Yang-Mills theories~\cite{Athenodorou:2021qvs}.}
    \label{tab:KSratios}
\end{table}

Provided the observed universality is not coincidental, the holographic glueball spectrum provides a rare opportunity to test non supersymmetry-protected sectors of holographic correspondence. It shows that despite the peculiarities of the holographic limit, first principle lattice calculations can be used to access physical observables in this regime.

 Note that the Klebanov-Strassler theory is not expected to give precisely the same results as the $SU(\infty)$ bosonic Yang-Mills. The results summarized in figure~\ref{fig:fullspec} is rather  the spectrum of the supersymmetric $SU(\infty)$ Yang-Mills, deformed by the presence of additional matter fields. The numerical consistency of the results in the KS theory and in the bosonic Yang-Mills suggests that the KS theory  would give a similar, or even better match with the spectrum of the supersymmetric pure Yang-Mills. The supersymmetric case would give us a larger base for comparison with the multitude of additional states, but unfortunately it is a much more challenging case to consider on the lattice. We hope that such a comparison will be possible in the future.

As a separate question we would like to discuss implications of holographic results  for the odderon physics. First subtlety is that glueball Regge trajectories are expected to have flatter slopes as compared to the universal slope of other hadrons. This is attributed to the structure of glueballs, which unlike mesons or baryons, are comprised  of  particles connected by a flux tube  in the adjoint representation. Alternatively glueballs can be seen as closed fundamental flux tubes.  Two different pictures indicate there is no universal glueball Regge trajectory slope, but in both cases the value is smaller than for other hadrons. These theoretical expectations are supported by the lattice studies of pure glue theories and more broadly by experimental studies of the $pp$ and $p\bar{p}$ cross sections. Lattice predicts the slope of the leading pomeron trajectory, of the $2^{++}$ glueball, to be $\alpha'_{\rm P}\sim 0.28\alpha'$~\cite{Meyer:2004jc}, consistent with the experimentally observed value~\cite{Jaroszkiewicz:1974ep}, where $\alpha'$ is the slope of the meson trajectory. The subleading trajectory of the lightest $0^{++}$ glueball state on the lattice  exhibits a larger slope than $\alpha'_{\rm P}$ but still smaller than $\alpha'$.

Now, if one assumes the leading odderon trajectory to have the same slope with the leading pomeron trajectory, than the intercepts of the two are related by
\be
j^- \ = \ 1 - (2-j^+)\left(\frac{m_{1^{--}}^2}{m_{2^{++}}^2}\right). 
\label{jj}
\ee
The intercept of the leading pomeron trajectory is known experimentally to be slightly above unity, $j^{+}\sim 1.08$~\cite{Donnachie:1992ny}. One could expect that the robustness of the mass ratios, through \eqref{jj}, can then provide a good estimate for $j^-$. However, this  relation predicts a rather low value for the intercept of the $1^{--}$ trajectory $j^{-}$, using lattice result for ${m_{1^{--}}^2}/{m_{2^{++}}^2}$ we find 
 $j^{-}\sim -1.3$ (for holographic models we find $j^{-}\sim -1.5$, which is consistent with the value obtained in early lattice estimates~\cite{Kaidalov:1999yd,Kaidalov:1999de}). Above we used the  mass ratios,   
\be
\frac{m_{1^{--}}}{m_{2^{++}}} \ \simeq \  1.58 ~\text{(lattice~\cite{Athenodorou:2021qvs})} \qquad \text{or} \qquad \frac{m_{1^{--}}}{m_{2^{++}}} \ \simeq \  1.64 ~\text{(holography~\cite{Dymarsky:2008wd})}\,.
\ee
Similar picture was outlined in \cite{Meyer:2004jc}, which used  lattice to study Regge  trajectories. If one assumes the $1^{--}$ state  to be on the same trajectory with $3^{--}$, then this trajectory's slope is numerically close to $\alpha'_{\rm P}$. At the same time, a perturbative analysis (hard odderon) predicts the intercept $j^-\sim 1$ and a slope smaller than $\alpha'_{P}$~\cite{Ewerz:2003xi}. Interpolation between strong and weak coupling was studied in~\cite{Brower:2014wha} in ${\cal N}=4$ supersymmetric Yang-Mills theory. It shows $j^{-}$ and $\alpha'_{\rm P}$ can vary substantially between weak and strong coupling, thus providing consistency to  the picture above.

A common experimental view, however, is that the intercept of the odderon trajectory is slightly below unity. Should $1^{--}$ belong to this trajectory, the slope of the latter would be very small compared to that of the pomeron. This is in sharp contrast to the picture above. Hence, this may indicate  that $1^{--}$ belongs to a subleading (rather than leading) trajectory. A possible scenario is that the leading trajectory could be that of the $3^{--}$ state, as pointed out in \cite{Meyer:2004jc}. This scenario though is beyond the scope of the holographic approximation discussed in this note. In other words, the structure of the leading odderon trajectory is an interesting open question, which will hopefully be addressed experimentally.

\paragraph{Acknowledgements.} The authors thank  I.~Klebanov, C.~Rodrigues Filho and C.~Royon for discussions. AD is supported by the NSF under grant PHY-2013812. The work of DM was done with support of the Simons Foundation, award \#884966, via the Association International Institute of Physics (AIIF) and of the grant of the agency CNPq of the Brazilian Ministry of Science, Technology and Innovation \#433935/2018-9.

 This note is an extended version of a contribution to proceedings of the Low X 2021 conference on Elba, based on a talk given at that event.

\bibliographystyle{JHEP}
\bibliography{refs}

\end{document}